\documentclass[reprint,3p,times,onecolumn,12pt]{elsarticle} 
\usepackage{amsmath,amsthm}
\usepackage{amsfonts}
\usepackage{amssymb}
\usepackage{subfigure}

\usepackage{hyperref}
\usepackage{graphicx}
\usepackage{latexsym}
\usepackage{amssymb}
\usepackage{amscd,amssymb}
\usepackage[arrow,matrix]{xy}
\usepackage{graphicx}
\usepackage{epstopdf}
\usepackage{subcaption}
\usepackage{soul}
\biboptions{sort&compress}
\usepackage{setspace}
\usepackage{amscd,amsmath,amsthm,amssymb,color}
\usepackage{xcolor}

\allowdisplaybreaks
\theoremstyle{plain}

\emergencystretch=\maxdimen
\newcommand{\bea}{\begin{eqnarray}}
\newcommand{\eea}{\end{eqnarray}}
\newcommand{\bes}{\begin{subequations}}
	\newcommand{\ees}{\end{subequations}}

\usepackage{xcolor}

\begin{document}
\title{Computing the Dynamics of Multi-Lumps in Nonlinearity-Managed Spatial-Symmetric Dispersive Wave Framework}

\author[gitam]{Sudhir Singh}
\address[gitam]{Department of Mathematics \& Statistics, School of Sciences, GITAM (Deemed to be University), \\ Bengaluru -- 561 203, Karnataka, India}

\author[vslp]{P. Tripathi}
\address[vslp]{School of Computational and Data Sciences, Vidyashilp University, \\ Bengaluru -- 562 110, Karnataka, India}

\author[trp,easw]{K. Manikandan}
\address[trp]{Center for Nonlinear and Complex Networks, SRM TRP Engineering College, \\ Tiruchirappalli -- 621 105, Tamil Nadu, India}
\address[easw]{Center for Research, Easwari Engineering College, Chennai -- 600 089, Tamil Nadu, India}

\author[pcm]{K. Sakkaravarthi}
\address[pcm]{Department of Physics, Pachaiyappa’s College for Men, Kanchipuram -- 631501, Tamil Nadu, India}

\cortext[cor]{{Email: sudhirew@gmail.com, ssingh8@gitam.edu (Sudhir Singh)\newline 
\indent \quad priyambada.t@vidyashilp.edu.in (P. Tripati)\newline 
\indent \quad 
manikandan.cnld@gmail.com (K. Manikandan)\newline 
\indent \quad ksakkaravarthi@gmail.com (K. Sakkaravarthi)}}
	
\journal{Applied Mathematics Letters}
	
\date{\today}

\setstretch{1.20}	
\begin{abstract} 
	We investigate the dynamics of multi-lump waves in a new version of a generalized spatial-symmetric higher-dimensional nonlinear dispersive water wave model using an analytical approach. This involves the proposition of a new spatial-symmetric nonlinear model in (3+1)-dimensions and the construction of its explicit solutions for multi-lump waves through a systematic analytical framework by employing Hirota's bilinear method and generalized polynomial expansions. Analyzing the resultant explicit solutions on their dynamical characteristics reveals that the obtained multi-lump waves are non-interacting and depict different geometrical patterns. The observed results demonstrate the significance of new higher-dimensional nonlinear dispersive models in enhancing our understanding of the dynamics of various types of localized waves.

\end{abstract}
	
\maketitle
	
\setstretch{1.250}	
\section{Introduction}
The manifestation of localized coherent structures in physical systems is one of the most intriguing phenomena in nature. The classical stable coherent structure, the solitons, appears in dispersive models with a delicate balance between nonlinearity and dispersion. A variety of one- and higher-dimensional models are proposed for the exploration of exotic wave patterns, such as solitary waves, breathers, cnoidal waves, compactons, rogue waves, and lumps, each characterized by distinct dynamical features and localization properties \cite{yang2010nonlinear}. 
Recently, spatially symmetric nonlinear models have garnered considerable interest, and several works have been reported in the literature for investigating localized wave structures, particularly solitons and lump wave solutions. To be specific, a (2+1)-dimensional spatially symmetric Hirota-Satsuma-Ito model has been analyzed for lump waves in \cite{ma2021nonlinearity}. Similarly, a (2+1)-dimensional dispersive water wave model with spatial symmetry is examined in \cite{ma2023lump}. The lump waves in the spatially symmetric KP model are reported in \cite{ma2023dispersion}, while a generalized spatially symmetric KP-type model is proposed in \cite{ma2024lump} with lump waves, breather solitons \cite{zou2024exact} and interaction solutions \cite{kang2025searching}. These works deal with first-order lump, interactions and breather wave solutions of (2+1)-dimensional models. 

Motivated by these developments and several open questions on this new avenue, our objective in this work is to propose a new spatial symmetric model and to analyze the dynamics of higher-order lump waves by constructing explicit solutions. For this purpose, we propose the following new generalized (3+1)-dimensional spatial symmetric nonlinear dispersive wave (GSSNDW) model: 
\begin{equation}
\begin{aligned}
P(u, v, w, r) 
&= \alpha_1 \left( u_{xt} + 6 u_x v_x + 6 u_{xx} v \right) 
   + \alpha_2 u_{xxxx} + \alpha_3 u_{xx} \\
&\quad + \beta_1 \left( u_{yt} + 6 u_y w_y + 6 u_{yy} w \right) 
   + \beta_2 u_{yyyy} + \beta_3 u_{yy} \\
&\quad + \gamma_1 \left( u_{zt} + 6 u_z r_z + 6 u_{zz} r \right) 
   + \gamma_2 u_{zzzz} + \gamma_3 u_{zz} = 0,
\end{aligned}
\label{eq1}
\end{equation}
where $v_y=u_x$, $w_x=u_y$ and $r_x=u_z$, with $u$ being a function of $x$, $y$, $z$ and $t$, while $\alpha_j$, $\beta_j$, and $\gamma_j$ with $j=1,2,3,$ are arbitrary real constants. The considered GSSNDW model reduces to a spatially symmetric (2+1) dimensional dispersive spatial symmetric model of KP type \cite{ma2023dispersion}. As mentioned above, our aim is to construct explicit solutions for multi-lump waves using Hirota's bilinear approach and to investigate their characteristic dynamics. 

The manuscript is structured as follows. In the next section, we construct first-order lump wave solutions by employing Hirota's bilinear method. Section 3 is devoted to the construction of higher-order lump wave solution with a detailed analysis of second- and third-order lump waves. Finally, concluding remarks are presented in the last section.

\section{The Method and First-order Lump-wave Solution}\label{sec:lin}
In this section, we study the evolutionary dynamics of the first-order lump wave of the generalized GSSNDW model (\ref{eq1}). To achieve this, we make use of the dimensional reduction and Hirota's bilinear methods. First, we transform the GSSNDW equation (\ref{eq1}) to a nonlinear $(1+1)$ dimensional equation through the scaling transformation $\xi =(  x - c t)$, $\eta = ( y +  z)$ and thereby apply the following set of logarithmic transformations:
\begin{equation} \label{btr}
u(\xi, \eta) = 2 (\ln \mathcal{F})_{ \xi \eta}, \quad v(\xi, \eta) = 2 (\ln \mathcal{F})_{\xi \xi}, \quad w(\xi, \eta) = 2 (\ln \mathcal{F})_{\eta \eta}, \quad r(\xi, \eta) = 2 (\ln \mathcal{F})_{\eta \eta}. 
\end{equation} 
From further simplification, we arrive at the Hirota bilinear equation described as
\begin{equation} \label{eq2.11}
\left( \alpha_1 D_{\xi}^4 + \chi_1 D_{\eta}^4 + \chi_2 D_{\xi}^2 + \chi_3 D_{\eta}^2 - c \chi_1  D_{\xi} D_{\eta} \right) \mathcal{F} \cdot \mathcal{F} = 0,
\end{equation} 
where $\chi_1 = ( \beta_1 + \gamma_1), \quad
\chi_2 = (\alpha_3-c \alpha_1),$ and $
\chi_3 = ( \beta_3 + \gamma_3 ).$ 
The above bilinearizing process possesses certain restrictions among parameters as
$\alpha_2 = \alpha_1$, 
$\beta_2 = \beta_1$ 
and $\gamma_2 = \gamma_1$. This shows that the model considered with nine parameters  (\ref{eq1}) reduces to the Hirota bilinear equation with six arbitrary parameters. 
In the above bilinear equation (\ref{eq2.11}), $D$ is the standard Hirota bilinear operator defined in Ref. \cite{hirota2004direct} and thereafter as followed by thousands of research works.

\begin{figure}[!ht]
    \centering
    \subfigure[$t = -45$]{
        \includegraphics[width=0.30\textwidth]{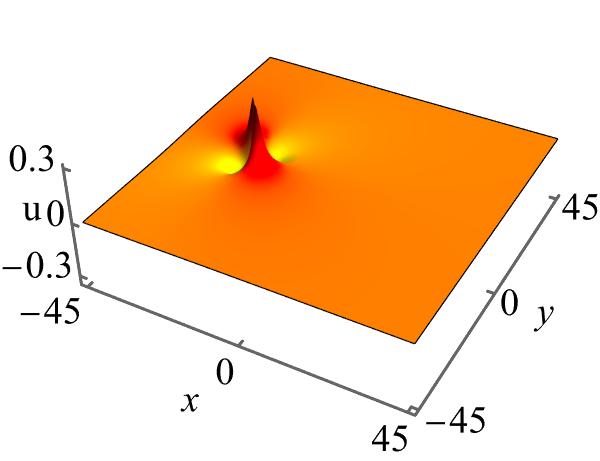}
        \label{}}
    \subfigure[$t = 0$]{
        \includegraphics[width=0.30\textwidth]{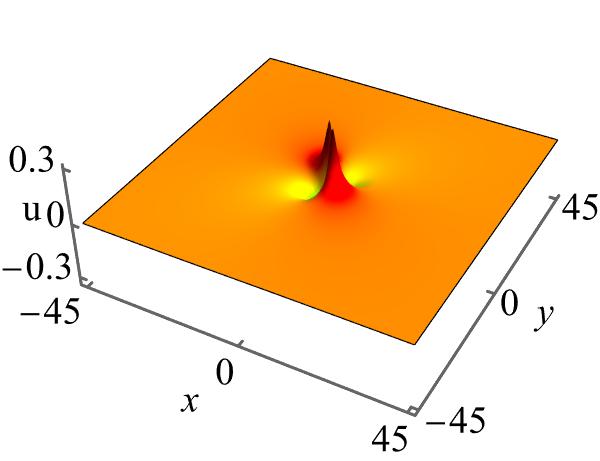}
        \label{figa22}
    }
    \subfigure[$t = 45$]{
        \includegraphics[width=0.30\textwidth]{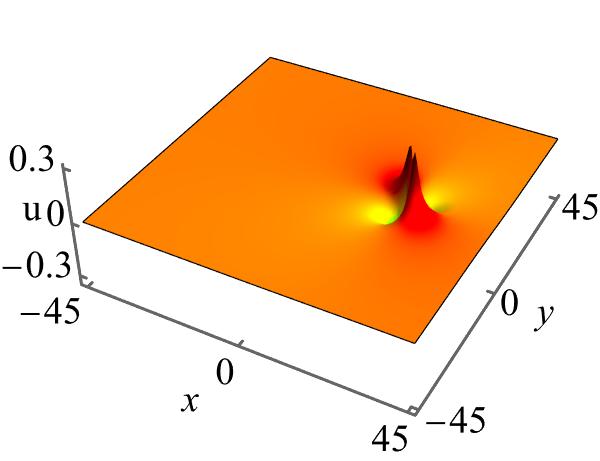}
        \label{figa3}
    }
        \label{figa6}
    \caption{Evolution of first-order lump  structure in the $x$--$y$ plane at $z = 0.1$ and other remaining parameters are taken as  $\alpha_1 = 3$, $\alpha_3 = -2.5$, $\beta_1 = -2$, $\beta_3 = -1$, $\gamma_1 = 3$, $\gamma_3 = -4$ and $c=0.5$. A similar profile evolution can be obtained along $x$--$z$ and $y$--$z$ planes at a fixed $y$ and $x$, respectively.}
    \label{fig:combined1}
\end{figure}

To construct the first-order lump-wave solution, we choose the following test function: 
\begin{equation}\label{eq4}
\mathcal{F}=F_1(\xi, \eta) = a_{0,0} + a_{2,0} \xi^2 + a_{0,2} \eta^2 + a_{1,1} \xi \eta,
\end{equation}
where $a_{0,0}$, $a_{2,0}$, $a_{0,2}$, and $a_{1,1}$ are constants to be determined. To proceed with the calculation, without loss of generality, we take $a_{0, 2}=1$.
By substituting the above first-order lump wave quadratic polynomial ansatz into the Hirota bilinear equation (\ref{eq2.11}) and solving it recursively, we get the following explicit form for the constant parameters:
\[
a_{0,0} = \frac{12  \left( \chi_1 \chi_2^2 + \alpha_1 \chi_3^2 \right)}{\chi_2 \left( c^2 \chi_1^2  - 4 \chi_2 \chi_3 \right)}, \quad
a_{2,0} = \frac{ \chi_3}{\chi_2}, \quad
a_{1,1} = \frac{c \chi_1}{\chi_2}.
\]
Substituting the above parameters into equation~(\ref{eq4}), applying the bilinearizing logarithmic transformation (\ref{btr}) and turning to the original model through the dimensional reduction, we obtain the explicit form for the required first-order lump wave solution. The resultant solution contains seven arbitrary parameters ($\alpha_1$, $\alpha_3$, $\beta_1$, $\beta_3$, $\gamma_1$, $\gamma_3$ and $c$), 
using which one can manipulate the dynamical characteristics of the obtained lump waves. The evolution of such a first-order lump wave is depicted in Fig. 
\ref{fig:combined1} at different times. 

It is evident from the above solution and Fig. \ref{fig:combined1} that the obtained first-order lump wave possesses paired peaks, and they travel in the three-dimensional space. 
By manipulating arbitrary parameters, the properties of the lump waves, such as amplitude, thickness/width, spacing between the pair of peaks, and their orientation, can be controlled effectively to demonstrate the desired effects. Considering the length of the manuscript, we have not provided further explanations and more graphical details. 

\section{Higher-order Lump Waves} 
To derive the higher-order lump wave solutions of the considered GSSNDW model~\eqref{eq1}, we incorporate technical input from the previous reports given in \cite{guo2021multiple, peng2021different}. By imposing the constraint conditions, we wish to eliminate the contribution of the mixed Hirota derivative \( D_{\xi}D_{\eta} \) and the fourth-order Hirota derivative with respect to the second variable \( D_{\eta}^4 \), thus simplifying the underlying model. 
Taking into account the constraint condition $\chi_1 = 0$, the restriction $\gamma_1 = -\beta_1$ have to be imposed.
To construct higher-order lump wave solutions, we take a generalized polynomial function for $\mathcal{F}$ by following Refs. \cite{clarkson2017rational, zhaqilao2018symbolic, guo2021multiple} as given below.
\begin{equation}\label{eq6}
\mathcal{F} =\mathcal{F}_{n+1}(\xi, \eta, \phi, \psi)= F_{n+1}(\xi, \eta) + 2 \phi \, \eta\, P_n(\xi, \eta) + 2 \psi \, \xi\, Q_n(\xi, \eta) + (\phi^2 + \psi^2)\, F_{n-1}(\xi, \eta),
\end{equation}
where
\[
F_n(\xi, \eta) = \sum_{k=0}^{n(n+1)/2} \sum_{i=0}^{k} a_{n(n+1)-2k,\, 2i}\, \eta^{2i}\, \xi^{n(n+1)-2k},
\]
\[
P_n(\xi, \eta) = \sum_{k=0}^{n(n+1)/2} \sum_{i=0}^{k} b_{n(n+1)-2k,\, 2i}\, \, \xi^{2i}\, \eta^{n(n+1)-2k},
\]
\[
Q_n(\xi, \eta) = \sum_{k=0}^{n(n+1)/2} \sum_{i=0}^{k} c_{n(n+1)-2k,\, 2i}\, \eta^{2i}\, \xi^{n(n+1)-2k}. 
\]
The application of the above polynomial form for $\mathcal{F} $ in the bilinear equation and solving the resulting form, one can arrive at the solution in (1+1)-dimensional version, from which the required solution for the (3+1)D GSSNDW equation can be constructed by using the dimensional-expansion.\\

To demonstrate the above solution process more clearly, we construct the second-order lump solution. For this purpose, we choose $a_{(n+1)(n+2),\,0} = 1$,  $F_0 = 1$, and for $n = 1$, which terminates the polynomial expansion for $\mathcal{F}$ as below.
\begin{equation}
    \mathcal{F}=\mathcal{F}_2(\xi, \eta, \phi, \psi) = F_2(\xi, \eta) + 2 \phi \, \eta\, P_1(\xi, \eta) + 2 \psi \, \xi\, Q_1(\xi, \eta) + (\phi^2 + \psi^2).
\end{equation}
The substitution of $F_2(\xi, \eta)$, $P_1(\xi, \eta)$, and $Q_1(\xi, \eta)$, gives the following detailed form for the test function $\mathcal{F}$ corresponding to the second-order lump wave: 
\[
\begin{aligned}\label{eq3}
\mathcal{F} =\;& (a_{0,0} + a_{0, 2} \eta^2 + a_{0, 4} \eta^4 + a_{0, 6} \eta^6) + (a_{2, 0} + a_{2, 2} \eta^2 + a_{2, 4} \eta^4) \xi^2 + (a_{4, 0} + a_{4, 2} \eta^2) \xi^4 + \xi^6 \\
&+ 2\phi \eta (b_{0, 0} + b_{0, 2} \eta^2 + b_{2, 0} \xi^2) + 2\psi \xi (c_{0, 0} + c_{0, 2} \eta^2 + c_{2, 0} \xi^2) + (\phi^2 + \psi^2) \, .
\end{aligned}
\]
Using the above test function~\eqref{eq3} in the bilinear equation~\eqref{eq2.11}, and collecting all terms arising at different powers of $\left( \xi^{i} \eta^{j} \right)$, where $\left( i, j = 0, 1, 2, \dots \right)$, we obtain a system of nonlinear algebraic equations. Upon solving this system of equations recursively, we arrive at the following parameters corresponding to the second-order lump solution:\\
\[
\left\{
\begin{aligned}
a_{0,0} &= -1875\left(\frac{\alpha_1}{\chi_2}\right)^3
\;+\; \frac{b_{2,0}^2\, \chi_3\, \phi^2}{9\, \chi_2}
\;+\; c_{2,0}^2\, \psi^2
\;-\; (\phi^2
\;+\; \psi^2)
 , \quad 
a_{0,2} = \frac{475\, \alpha_1^2}{\chi_2\, \chi_3} \,,  ~a_{0,4} = -\frac{17\, \alpha_1\, \chi_2}{\chi_3^2} \, ; \\
a_{0,6} &=\left( \frac{\chi_2}{\chi_3}\right)^3 \, , \quad 
a_{2,0} = -125 \left(\frac{\, \alpha_1}{\chi_2}\right)^2 \, , \quad 
a_{2,2} = -\frac{90\, \alpha_1}{\chi_3} \, , \quad 
a_{2,4} = 3\left(\frac{\, \chi_2}{\chi_3}\right)^2 \, ; \\
a_{4,0} &= -\frac{25\, \alpha_1}{\chi_2} \, , \quad 
a_{4,2} = \frac{3\, \chi_2}{\chi_3} \, , \quad 
b_{0,0} = -\frac{5\, b_{2,0}\, \alpha_1}{3\, \chi_2} \, , \quad 
b_{0,2} = -\frac{b_{2,0}\, \chi_2}{3\, \chi_3} \, ; \\
c_{0,0} &= \frac{c_{2,0}\, \alpha_1}{\chi_2} \, , \quad 
c_{0,2} = -\frac{3\, c_{2,0}\, \chi_2}{\chi_3} \,.
\end{aligned}
\right.
\]

\begin{figure}[!ht]
    \centering
    \subfigure[$t = -45$]{
        \includegraphics[width=0.30\textwidth]{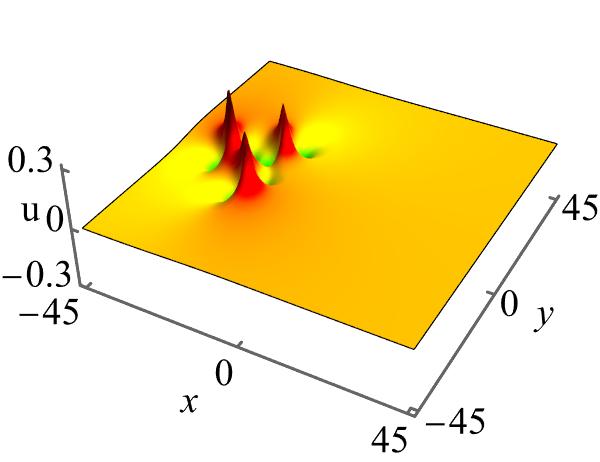}
        \label{figa33}
    }
    \subfigure[$t = 0$]{
        \includegraphics[width=0.30\textwidth]{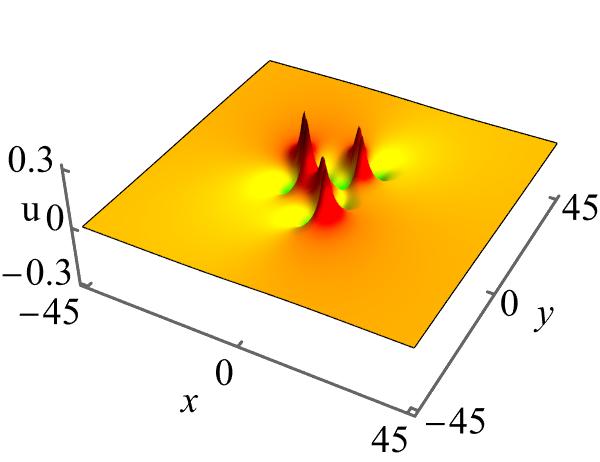}
        \label{figa27}
    }
    \subfigure[$t = 45$]{
        \includegraphics[width=0.30\textwidth]{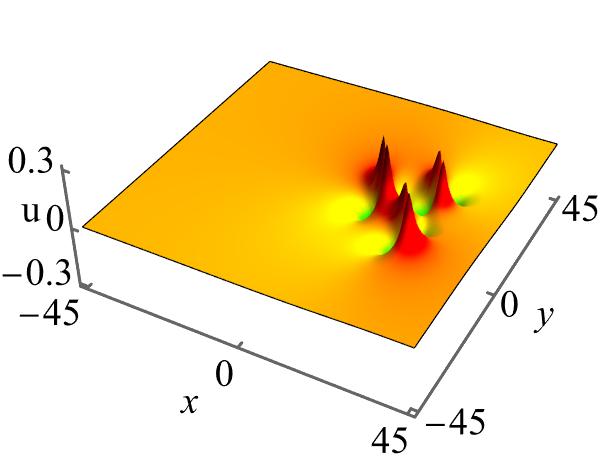}
        \label{figa3b}
    }
    \caption{The evolution of second-order lump-wave possessing triangular structure in the $x$--$y$ planes for $z = 0.1$  with other parameters taken as $\alpha_{1} = 5,\;
\alpha_{3} = -2.5,\;
\beta_{3} = -4.5,\;
\gamma_{3} = -4,\;
c = 0.5,\;
\phi = -3,\;
\psi = 5,\;$ and $
b_{2,0}=c_{2,0}  = 100.$}

    \label{fig:combined2}
\end{figure}

The above explicit form of parameters and the final expression $\mathcal{F}$ gives the required second-order lump-wave solution of the GSSNDW equation (\ref{eq1}) through the bilinear transformations and mapping to the original (3+1)-dimensions. 
This second-order solution contains four arbitrary solution parameters, namely, $b_{2, 0}, c_{2, 0}, \phi$, and $\psi$ apart from the arbitrary model parameters, and using them all, one can modify the nature and evolution of the lump-wave dynamics effectively. 
For an easy understanding, one of these evolutions of second-order lump waves is demonstrated in Fig. \ref{fig:combined2}. It is clear that the second-order lumps exhibit three sets of paired peaks and do not interact during the time evolution. Such a structure resembles the identities of three independent first-order lump-waves, and they combine to form a triangular geometry. As mentioned earlier, the characteristics of the triangular lump waves, like inter- and intra-peak distance, peak amplitude, width of peaks, angle or orientation of the peaks in addition to the orientation of the triangular structure itself, can be tuned by properly adjusting the available arbitrary parameters. 

To proceed further for the construction of a third-order lump-wave solution, we terminate the polynomial expansion for \( n = 2 \) and obtain the following form for the function $\mathcal{F}$, which will serve as the test function for the third-order lump solution:
\[
\mathcal{F}=\mathcal{F}_3(\xi, \eta, \phi, \psi)  = F_3(\xi, \eta) + 2 \phi \, \eta\, P_2(\xi, \eta) + 2 \psi \, \xi\, Q_2(\xi, \eta) + (\phi^2 + \psi^2) F_1(\xi, \eta) .
\]
where $F_1$ is given in $(\ref{eq4})$. By substituting the above $f(\xi, \eta)$ into the bilinear equation and solving the resulting set of algebraic equations recursively, as performed in the previous cases, one can arrive at the explicit form of the parameters as follows. 
\[
\left\{
\begin{aligned}
a_{0,0}  &= \frac{\alpha_{1} (878826025 \alpha_{1}^{5} \chi_{2}^{2} + 27 \phi^{2} \chi_{2}^{6} \chi_{3} - 27 \phi^{2} \chi_{3}^{7} - 27 \chi_{2}^{7} \psi^{2} + 27 \chi_{2}^{6} \chi_{3} \psi^{2})}{9 \chi_{2}^{8}}; \\
a_{0,2}  &= \frac{-300896750 \alpha_{1}^{5} \chi_{2}^{2} - 3 \phi^{2} \chi_{2}^{6} \chi_{3} + 3 \phi^{2} \chi_{3}^{7} + 3 \chi_{2}^{7} \psi^{2} - 3 \chi_{2}^{6} \chi_{3} \psi^{2}}{3 \chi_{2}^{6} \chi_{3}}; \\
a_{0,4}  &= \frac{16391725 \alpha_{1}^{4}}{3 \chi_{2}^{2} \chi_{3}^{2}}, \quad
a_{0,6}  =  \frac{-798980 \alpha_{1}^{3}}{3\chi_{3}^{3}}, \quad
a_{0,8}  =  \frac{4335 \alpha_{1}^{2} \chi_{2}^{2}}{\chi_{3}^{4}}, \quad
a_{0,10} =  \frac{-58 \alpha_{1} \chi_{2}^{4}}{\chi_{3}^{5}};  \\
a_{0,12} &=  \frac{\chi_{2}^{6}}{\chi_{3}^{6}},  \quad a_{2,0}  = \frac{-159786550 \alpha_{1}^{5} \chi_{2}^{2} - 3 \phi^{2} \chi_{2}^{6} \chi_{3} + 3 \phi^{2} \chi_{3}^{7} + 3 \chi_{2}^{7} \psi^{2} - 3 \chi_{2}^{6} \chi_{3} \psi^{2}}{{3 \chi_{2}^{7}} }; \\
a_{2,2}  &= \frac{565950 \alpha_{1}^{4}}{\chi_{2}^{3} \chi_{3}}, \quad
a_{2,4}  =  \frac{14700 \alpha_{1}^{3}}{\chi_{2} \chi_{3}^{2}}, \quad
a_{2,6}  =  \frac{35420 \alpha_{1}^{2}\chi_{2}}{\chi_{3}^{3}}, \quad
a_{2,8}  =  \frac{-570 \alpha_{1} \chi_{2}^{3}}{\chi_{3}^{4}}, \quad
a_{2,10} =  \frac{6 \chi_{2}^{5}}{\chi_{3}^{5}}; \\
a_{4,0}  &= \frac{-5187875 \alpha_{1}^{4}}{3 \chi_{2}^{4}}, \quad
a_{4,2}  =  \frac{-220500 \alpha_{1}^{3}}{\chi_{2}^{2} \chi_{3}}, \quad
a_{4,4}  =  \frac{37450 \alpha_{1}^{2}}{\chi_{3}^{2}}, \quad
a_{4,6}  =  \frac{-1460 \alpha_{1} \chi_{2}^{2}}{\chi_{3}^{3}}, \quad a_{4,8}  = \frac{15 \chi_{2}^{4}}{\chi_{3}^{4}};
\end{aligned}
\right.
\]
\[
\left\{
\begin{aligned}
a_{6,0} &=  \frac{-75460 \alpha_{1}^{3}}{3\chi_{2}^{3}}, \quad
a_{6,2} =  \frac{18620 \alpha_{1}^{2}}{\chi_{2}  \chi_{3}}, \quad
a_{6,4}  =  \frac{-1540 \alpha_{1}\chi_{2}}{\chi_{3}^{2}}, \quad
a_{6,6}  =  \frac{20 {\chi_{2}^{3}}}{\chi_{3}^{3}}, \quad a_{8,0}  =  \frac{735 \alpha_{1}^{2}}{\chi_{2}^{2}};  \\ 
a_{8,2}  &=  \frac{-690 \alpha_{1}}{\chi_{3}}, \quad
a_{8,4}  =  \frac{15\chi_{2}^{2}}{\chi_{3}^{2}}, \quad
a_{10,0} =  \frac{-98 \alpha_{1}}{\chi_{2}}, \quad
a_{10,2} =  \frac{6 \chi_{2}}{\chi_{3}}, \quad b_{0,0}  = \frac{-18865 \alpha_{1}^{3} \chi_{3}^{3}}{3 \chi_{2}^{6}};  \\
b_{0,2}  &=  \frac{-665 \alpha_{1}^{2} \chi_{3}^{3}}{\chi_{2}^{5}}, \quad
b_{0,4}  =  \frac{-105 \alpha_{1} \chi_{3}^{3}}{\chi_{2}^{4}}, \quad
b_{0, 6}  =  \frac{5 \chi_{3}^{3}}{\chi_{2}^{3}}, \quad
b_{2,0}  =  \frac{-245 \alpha_{1}^{2} \chi_{3}^{2}}{\chi_{2}^{4}}, \quad b_{2,2}  =  \frac{190 \alpha_{1} \chi_{3}^{2}}{\chi_{2}^{3}}; \\
b_{2,4}  &= \frac{-5 \chi_{3}^{2}}{\chi_{2}^{2}}, \quad
b_{4,0}  =  \frac{7 \alpha_{1}  \chi_{3}}{\chi_{2}^{2} }, \quad
b_{4,2}  =  \frac{-9 \chi_{3} }{\chi_{2}}, \quad
c_{0,0}  =  \frac{-12005 \alpha_{1}^{3}}{3\chi_{2}^{3}}, \quad
c_{0,2}  =  \frac{535 \alpha_{1}^{2}}{\chi_{2} \chi_{3}}, \quad c_{0,4}  =  \frac{-45 \alpha_{1} \chi_{2}}{\chi_{3}^{2}}; \\
c_{0,6}  &= \frac{5 \chi_{2}^{3}}{\chi_{3}^{3}}, \quad c_{2,0}  = \frac{-245 \alpha_{1}^{2}}{\chi_{2}^{2}}, \quad c_{2, 2}  = \frac{230 \alpha_{1}}{\chi_{3}}, \quad c_{2, 4}  = \frac{-5 \chi_{2}^{2}}{\chi_{3}^{2}}, \quad c_{4, 0}  = \frac{-13 \alpha_{1}}{\chi_{2}}, \quad c_{4, 2}  = \frac{-9 \chi_{2}}{\chi_{3}}.
\end{aligned}
\right.
\]

\begin{figure}[!ht]
    \centering
    \subfigure[$t=-75$]{
        \includegraphics[width=0.30\textwidth]{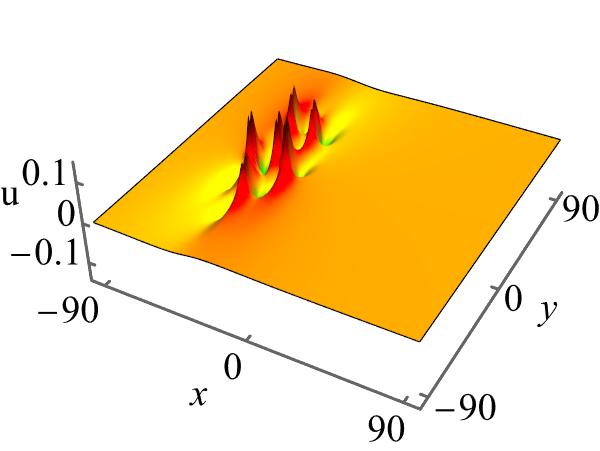}
        \label{figa1}
    }
    \subfigure[$t=0$]{
        \includegraphics[width=0.30\textwidth]{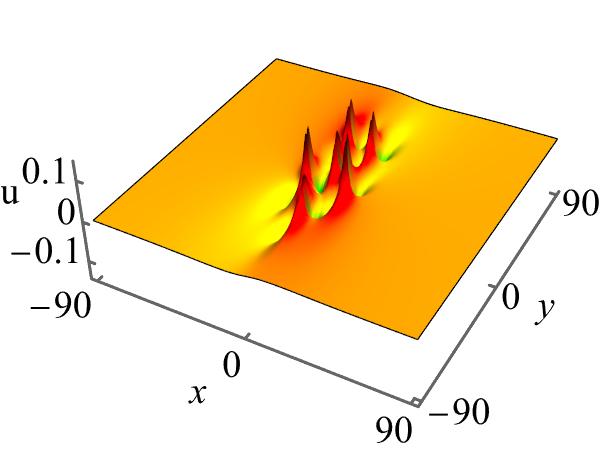}
        \label{figa2wz}
    }
    \subfigure[$t=75$]{
        \includegraphics[width=0.30\textwidth]{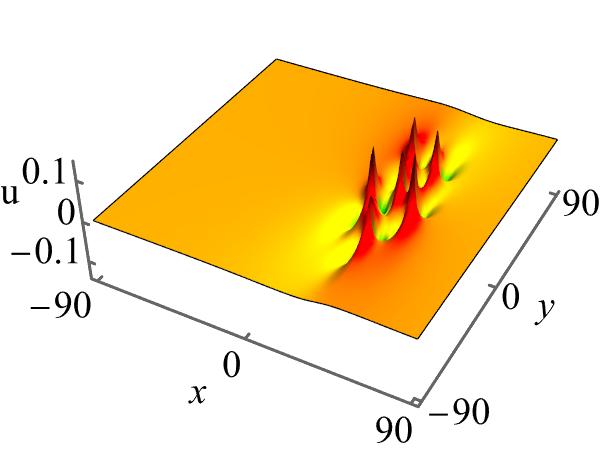}
        \label{figa3z}
    }
    \caption{The evolution of third-order lump-wave solution exhibiting a pentagon structure in the $x$--$y$ planes for $z = 0.1$  with other parameters as 
$\alpha_{1} = 0.2,\;
\alpha_{3} = -0.1,\;
\beta_{3} = -1.5,\;
\gamma_{3} = -0.15,\;
c = 0.5,\;$ and $
\phi = 200,\;
\psi = 100.$}
    \label{fig:combined3}
\end{figure}

Similar to the previous two cases, one can deduce the explicit form of the required solution by using the above parameter relations, bilinear transformations, and the dimensional expansions to the original (3+1)D GSSNDW model (\ref{eq1}). Furthermore, by adjusting the arbitrary parameters in the final solution, one can control the characteristics of the resulting lump waves. A clear analysis reveals that the present third-order lump-wave solution comprises six sets of paired peaks, which exhibit a pentagonal shape with five paired peaks on the edges and a central paired peak. Further, such a pentagon structure moves over time by maintaining the geometrical structure, and every property can be controlled by using the arbitrary parameters as one wishes to have. For completeness, we have shown an example of the resulting third-order lump wave dynamics in Fig. \ref{fig:combined3}. 

Based on the above analysis, the present procedure can be extended to construct the explicit form of the higher-order lump wave solution of arbitrary $n$-th order using the current generalized polynomial function, and their evolutionary dynamics can be explored. 
The present analysis exposes the formation of different geometrical structures at various orders of the lump wave solutions, which resemble the studies reported in the literature for standard and extended KP-type nonlinear equations. In future studies, other types of localized wave solutions, such as solitons, breathers, etc., can be constructed, and their characteristic dynamics can be explored in detail with possible applications. 

\section{Conclusions} \label{sec:perperturb}
To summarize the present work, we have investigated an important problem that showcases the dynamics of higher-order lump waves in a generalized (3+1)-dimensional nonlinearity-managed spatially symmetric dispersive model. By employing the Hirota bilinear method in conjunction with generalized polynomial expansions, we have constructed explicit form of first-, second-, and third-order lump-wave solutions and presented the solution methodology for an arbitrary $n$-th order solutions. 
Our analysis of the obtained solutions reveals that the lump waves possess paired peaks with non-interacting behavior and exhibit different geometrical patterns, namely simple dot, triangular, and pentagonal structures, with single, triple, and sextuple paired peaks, allowing for manipulation possibilities using the available arbitrary parameters. The observations can provide valuable insights into the dynamics of higher-order lump structures in any higher-dimensional water wave models and other diverse physical systems.

\setstretch{1.0}	
\subsection*{Acknowledgement} SS gratefully acknowledges the financial support from GITAM University through the GITAM New-Faculty Seed Grant (G-NSG 2025/0047). KM would like to thank SRM TRP Engineering College, India, for their financial support (Grant No. SRM/TRP/RI/005). KM is also supported by the Anusandhan National Research Foundation (ANRF), India — formerly known as the Science and Engineering Research Board (SERB), Government of India — through the MATRICS Research Grant (No. MTR/2023/000921). 

\setstretch{1.0}	
\bibliographystyle{acm}
\bibliography{stability.bib}
\end{document}